\documentclass[preprint,12pt]{elsarticle}

\usepackage{amsmath,amsfonts,amssymb}
\usepackage{lineno}

\begin{document}
\newcommand{\frc}[2]{\raisebox{2pt}{$#1$}\big/\raisebox{-3pt}{$#2$}}
\newcommand{\vect}[1]{{\textbf{\textit{#1}}}} 
\begin{frontmatter}
\title{Sommerfeld half-space problem revisited: Short-wave asymptotic solutions}
%
%
%

\author{Seil~Sautbekov}

\begin{abstract}
A method for solving the half-space Sommerfeld problem is proposed, which allows us to obtain exact solutions in the form of Sommerfeld integrals, as well as their short-wave asymptotics. The first carried out by reducing the Sommerfeld problem to solving a system of equations to surface current densities on an interface of media in the Fourier transform domain. The second is provided by the modified saddle point method using an etalon integral. Careful attention pays the integration technique of the Sommerfeld integrals.
The uniformly regular for any observation angles expressions of all types of waves, such as space, surface, and lateral, are obtained.  The simple asymptotic expressions for the reflected and transmitted space waves are found via a source field, bearing in mind the Fresnel coefficients. The asymptotic expressions for the space waves checked by the law of conservation of energy flux. The issue of the existence of surface waves analyzed in detail. The analysis and physical interpretation of the obtained expressions, as well as the applicability conditions, are carried out. The Sommerfeld integrals evaluated in closed form.
\end{abstract}

\end{frontmatter}

%

\section{Introduction}
%
%
%
%


\label{sec:introduction}
 
The famous half-space problem was first rigorously solved by Sommerfeld \cite{Sommerfeld1909}  a century ago when Sommerfeld first formulated the problem of the radiating vertical electric dipole over the interface by invoking the Hertz vector rather than the electromagnetic fields. With the use of the Fourier-Bessel representations in cylindrical coordinates, Sommerfeld proposed the solution for Hertz vector in terms of improper integrals now known as Sommerfeld integrals. Sommerfeld further derived an asymptotic surface wave expression, assuming high media contrast and large horizontal range.  The theory of surface waves was worked out by Zenneck \cite{Zenneck} and Sommerfeld \cite{Sommerfeld1909,Sommerfeld1926}. 
 
The propagation of radio waves, generated by a vertical  electric dipole located near the planar interface between two homogeneous and isotropic half spaces, such as the air over the surface of the earth or seawater, along the surface of the ground  is important in radio communication \cite{Fock1945,SWanten,Michalski}. 

The surface electromagnetic waves are of practical interest because their energy decreases inversely as the range from a point source, while the energy of space electromagnetic waves decreases inversely as the distance squared to the source. In practical work this factor may substantially extend the range of action of radars and communication systems and also increase their efficiency \cite{experimental}. 

It should be noted that in recent years there has been a resurgence of interest in the Sommerfeld half-space problem and the surface wave in particular, in the context of THz applications, near-field optics, plasmonics, and nanophotonics \cite{Jeon,Sih,Michal,Huang2016}.
Results of experiments for the observation of surface electromagnetic waves excited by a vertical dipole and propagating above the ice-coated surface of a salt lake \cite{Bash}, as well as in the THz spectral range generated on dense aluminum films covering the optical quality glass plates  \cite{Bogomol} are considered. The replication of ‘crucial’ Seneca Lake experiment of 1936 was conducted  2014 on the west shore of Seneca Lake \cite{corum}.
 
Accurate evaluation of the influence of the finite ground conductivity presents a practical concern in broad variety of applications.  Many aspects of excitation and propagation of the surface waves still remain uninvestigated. While Zenneck’s field solution is exact, it was considered to be false for many years. This problem has been discussed from a theoretical point of view for many years. For instance, although Weyl's approach \cite{Weyl} leads to  solution, which is equivalent to Sommerfeld's integral formula but does not contain a surface wave, because the steepest descent path never meets the pole corresponding to the  Zenneck wave and it is not removed as a residue from the Sommerfeld integral.  

Among other issues, this fact was the subject of discussion between two well-known physicists, H.M. Barlow and J.R. Wait \cite{Wait}. Wait expressed doubts that the Zenneck surface wave may exist in practical situations.  However, the experimental data \cite{Bash}, conducted in 2009, fully correspond to the Sommerfeld result and confirm that the field imitating Zenneck surface wave is absent with increasing the numerical distance within its limits.  In 1979 Hill and Wait \cite{Hill,wait1981} analytically found an aperture distribution that excites a pure Zenneck surface wave with no radiation field.    
Unfortunately, Zenneck wave has been surrounded by the controversies pertaining to their physical existence \cite{Schelkun,Sarkar,Sarkar14}. 
A historical account and extensive list of references can be found in \cite{Wait1998,Banos}.

It turns out that all particular cases of the problem of propagation of electromagnetic waves through a planar interface between two homogeneous media, usually solved by different methods \cite{Fock1946,Brekh,Ott42,Leontovic1946}, can be considered from one point of view.

This work is a continuation of a series of papers \cite{sautbek10,sautbek13,sautbek14,sautbek17,sautbek18} devoted to a novel method for solving the classical Sommerfeld problem of a vertical electric dipole radiating in an imperfectly conducting half-space. Where occurs the quantitative estimation carried out by various numerical methods of the short-wave asymptotic solutions, obtained by the steepest descent or saddle point method (SPM) using an etalon integral concerning, in the main, the reflected and surface waves.
   
In this paper, we reexamine the classical Sommerfeld half-space problem of a point vertical dipole. Since the method proposed earlier allows us to obtain rigorous solutions of Sommerfeld problem directly for electromagnetic field in an integral form \cite{sautbek10,sautbek18}, another outstanding issue is the choice of a more efficient calculation technique for Sommerfeld integrals, taking into account the singularities of the integrand. Such a integration technique, which is a modification of the SPM using the etalon integrals, was proposed in \cite{sautbek18} for calculating reflected and near-surface waves at a plane interface between two media. 

The purpose of this work is to obtain in closed form the simple short-wave asymptotic formulae for fields with conditions of applicability, with emphasis on their physical meaning for reflected, transmitted, lateral, and near-surface waves excited by surface currents at a planar interface of media, by means of SPM with using a special function for calculating the corresponding Sommerfeld integrals. 

In what follows, Section II recaps the fundamentals expressions derived in the previous work \cite{sautbek10} for the electromagnetic field in the spectral domain.
All asymptotic expressions for all types of waves scattered by a planar interface are given in Section III. 
Important findings are summarized in Sections IV and V. Derivations for most important statements and arguments are given in the appendices. The theoretical development is interpreted from a geometric-optics viewpoint and validated by the energy flux conservation law.

\section{The electromagnetic field integral expressions in spectral domain}
\label{2sec:state}

\begin{figure}[h]
\centering\includegraphics[width=3.2in]{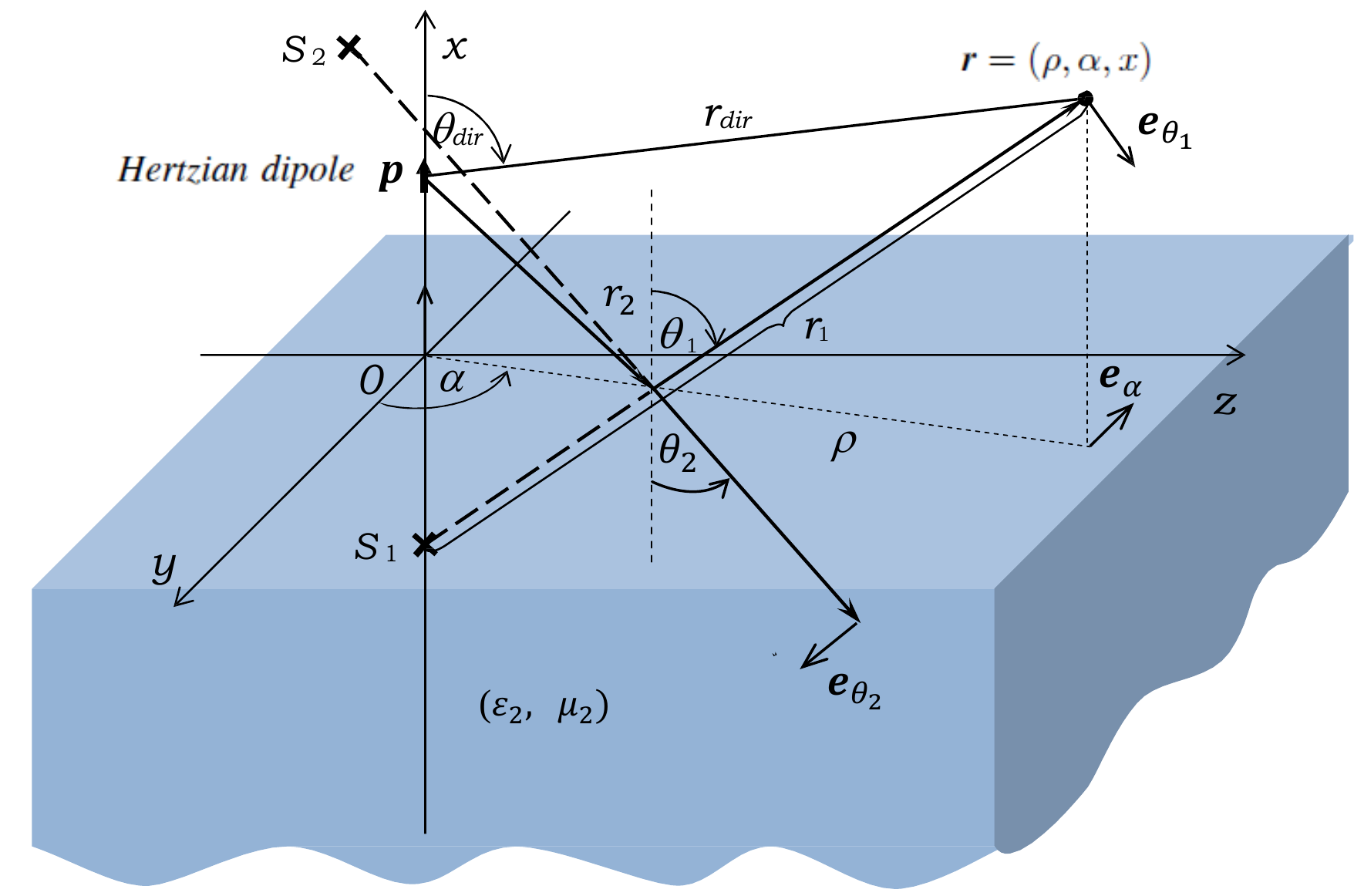}
\caption{Hertzian dipole with the moment  \vect p above a half-space medium.}\label{fig1}
\end{figure}

 A vertical point (Hertzian) dipole, characterized by dipole moment 
$\vect{p}=p\vect{e}_x$,
 $p=\text{const}$,
 is directed to the positive $x$ axis, at altitude $x_0$
   above infinite and flat ground. The problem geometry is provided in  Fig.~\ref{fig1}.  The dipole radiates time-harmonic electromagnetic (EM) waves at angular frequency 
   $\omega$ ($e^{-i\omega t}$ time dependence is assumed). The relative complex permittivity of the ground is:
 \begin{equation}\label{sig}
    \dot{\varepsilon}_2 = \varepsilon_2 + i~ \frac\sigma {\varepsilon_{0}\omega},
 \end{equation}     
where $\sigma$ is the ground conductivity, and $\varepsilon_{0}$ is the permitivity in vacuum. 

For better clarity of the calculation technique of the short-wave asymptotic behavior of fields, we present below the rigorous solutions in an integral form in a spherical coordinate system.

\subsection*{\textit{Direct Field}}
\label{subsec:dir_field_int}

The vertical electric dipole field in an integral form is given by \cite{sautbek18}
\begin{eqnarray}\label{Elos}
\vect{E}^{LOS}=-\frac{i p k_{01}}{8\pi\varepsilon_{0}\varepsilon_{1}}\int\limits_{-\infty}^{+\infty}
\vect{e}_{\theta}\left(k_{\rho}\right)\dfrac{k_{\rho}|k_{\rho}|}{\varkappa_{1}}\textrm{H}_{0}^{(1)}(k_{\rho}\rho)\nonumber\\e^{i\varkappa_{1}|x-x_{0}|}~dk_{\rho},\quad\varkappa_{1} = \sqrt{k_{01}^{2}-k_{\rho}^{2}},\\
 \vect{e}_{\theta}(k_{\rho})=-\vect{e}_{x}\,\frc{|k_{\rho}|}{k_{01}}+\vect{e}_{\rho}\,\textrm{sgn}(x-x_{0})\,\frc{\varkappa_{1}}{k_{01}}.\nonumber
\end{eqnarray}

 Similarly, the magnetic field 
\begin{equation}\label{eq:8}
\vect{H}^{LOS}=-\vect{e}_{a}\frac{i\omega p}{8\pi}\int_{-\infty}^{+\infty}\frac{|k_{\rho}|k_{\rho}}{\varkappa_{1}}\text{H}_{0}^{(1)}\left(k_{\rho}\rho\right)e^{i\varkappa_{1}|x-x_{0}|}~dk_{\rho}.
\end{equation}

\subsection*{\textit{Scattered Fields}}
\label{subsec:scat_field_int}

For the Reflected (index 'R') and Refracted or Transmitted (index 'T') electric and magnetic field respectively
  \cite{sautbek18}
\begin{eqnarray}\label{eq:ER}
\vect{E}^{R}=-\frac{k_{01}}{8\pi\varepsilon_{0}\varepsilon_{1}\omega}
\int\limits_{-\infty}^{+\infty}\vect{e}_{\theta_1}(k_{\rho})\tilde{J}^R\textrm{H}_{0}^{(1)}(k_{\rho}\rho)e^{i\varkappa_{1}x}k_{\rho}dk_{\rho},
\\\label{eq:10}
\vect{H}^{R}=-\frac{\vect{e}_a}{8\pi}\int\limits_{-\infty}^{+\infty}\tilde{J}^R
\text{H}_{0}^{(1)}\left(k_{\rho}\rho\right)
e^{i\varkappa_{1}x}k_{\rho}dk_{\rho},\phantom{ooooooojooo}
\\\label{eq:11}
\vect{E}^{T}=\frac{k_{02}}{8\pi\varepsilon_{0}\dot{\varepsilon}_{2}\omega}
\int\limits_{-\infty}^{+\infty}\vect{e}_{\theta_2}(k_\rho) \tilde{J}^T\text{H}_{0}^{(1)}\left(k_{\rho}\rho
\right)e^{-i\varkappa_{2}x}k_{\rho}dk_{\rho},\\
\label{eq:12}
\vect{H}^{T}=\frac{\vect{e}_a}{8\pi}\int\limits_{-\infty}^{+\infty}\tilde{J}^T
\text{H}_{0}^{(1)}\left(k_{\rho}\rho\right)e^{-i\varkappa_{2}x}k_{\rho}dk_{\rho},\phantom{ooooojjjooo} \\
\varkappa_2 = \sqrt{k_{02}^2-k_\rho^2},\quad \vect{e}_{\theta_1}(k_{\rho})=-\vect{e}_{x}\,\frc{|k_{\rho}|\footnotemark\nonumber}
 {k_{01}}+\vect{e}_{\rho}\,\frc{\varkappa_{1}}{k_{01}},\\
  \vect{e}_{\theta_2}(k_{\rho})=-\vect{e}_{x}\,\frc{|k_{\rho}|}{k_{02}}-\vect{e}_{\rho}\,\frc{\varkappa_{2}}{k_{02}},\phantom{o}
 \end{eqnarray}  
\footnotetext
{
The modulus of $|k_{\rho}|$ on the real axis can be represented as a function $\sqrt{k_\rho^2-(i\delta)^2}$, $0<\delta\to 0$ and it can be analytically extended to the whole complex plane so that the condition $\text{Re}\,k_\rho > 0$ is everywhere satisfied on one of the Riemann sheets, and so the cuts should be pass along the imaginary semi-axes outgoing from the branch points $\pm i\delta$, according to the equation $\textrm{Re}\sqrt{k_\rho^2+\delta^2}=0$.
Finally,
$|k_\rho| = k_\rho \,\textrm{sgn}(\textrm{Re}\,k_\rho)$ occurs for complex argument. It is useful to note that the contribution of the integral along the banks of the above-mentioned cuts is absent due to a factor $k_\rho |k_\rho|$.
 } 
\begin{equation} \label{eq:14}
\begin{cases}
& \tilde{J}^R = 
i\dfrac{\omega p|k_{\rho}|}{\varkappa_1}e^{i\varkappa_1x_0}R_{\parallel},
\\
&\tilde{J}^T =-i\dfrac{\dot{\varepsilon}_2 k_{01}}{\varepsilon_1k_{02}}\dfrac{\omega p|k_{\rho}|}{\varkappa_1}e^{i\varkappa_1x_0}T_\parallel
\end{cases}
\end{equation}
are the surface current densities $\tilde{J}^R $ and $\tilde{J}^T$, just above and below the interface level and act as the secondary sources for the reflected ($x > 0$) and refracted fields ($x < 0$) respectively.
Here $R_\parallel$, $T_\parallel$ are the Fresnel reflection and refraction coefficients:
\begin{equation}\label{RT}
R_\parallel=\frac{\dot{\varepsilon}_2\varkappa_1-
\varepsilon_1\varkappa_2}{\dot{\varepsilon}_2\varkappa_1+
\varepsilon_1\varkappa_2}, \quad 
T_\parallel=\frac{2\varepsilon_1\varkappa_1k_{02}}{k_{01}(\varepsilon_1\varkappa_2+\dot{\varepsilon}_2\varkappa_1)}
\end{equation}
since the expressions for the fields in (\ref{eq:ER}-\ref{eq:12}), in fact, represent the expansion of the electromagnetic field in term of inhomogeneous plane waves.

\section{Asymptotic Expressions for the EM Waves}
\label{sec:asym_expr}
Below we give short-wave asymptotic expressions for fields which derive from the integral representations (\ref{Elos}), (\ref{eq:ER}), (\ref{eq:11}). Integration is performed using SPM.

\subsection*{\textit{Direct Field}}

The source fields (Fig.\ref{fig1}) of the far-field zone can easily be obtained by the saddle-point technique from (\ref{Elos}) и (\ref{eq:8})  
\begin{eqnarray}\label{Alos}
\vect{E}^{LOS}(r_{dir},\theta_{dir})=- \vect{e}_{\theta}(\theta_{dir})\,\dfrac{pk_{01}^{2}}{4\pi\varepsilon_{0}\varepsilon_{1}}\dfrac{e^{i\varPhi(\theta_{dir})}}{r_{dir}}\sin\theta_{dir},\\
\label{eq:18}
\vect{H}^{LOS} (r_{dir},\theta_{dir})= -\vect{e}_a\dfrac{1}{Z_{1}}E^{LOS}(r_{dir},\theta_{dir}), \phantom{oooooooo}
\end{eqnarray}
where $\varPhi(\theta)=k_{01}r_{dir}\cos(\theta-\theta_{dir})$, $r_{dir}=\sqrt{(x-x_0)^2+\rho^2}$, $\sin\theta_{dir}=\rho/r_{dir}$, $Z_1=\sqrt{\frc{\mu_0\mu_1}{\varepsilon_0\varepsilon_1}}$, the wave impedance of the first medium.

\subsection*{\textit{Reflected Field}}
\label{subsec:scat_field_anal}

We will need the regular for any values of the observation angles $\theta_1$ asymptotic expressions  for reflected and transmitted waves obtained by SPM using the special function $X$ (\ref{xi}) \cite{sautbek18}
\begin{eqnarray}\label{ter}
\nonumber \vect{E}^{R}(r_1, \theta_1)=-\vect{e}_{\theta_{1}}\dfrac{pk_{01}^{3}}{2\varepsilon_{0}\varepsilon_{1}}\sqrt{\dfrac{-2i}{\pi k_{01}\rho}} e^{ik_{01}r_1\cos(\theta_{p}-\theta_1)}\phantom{O}\\
\sin^{\frac{3}{2}}\theta_{1}
\sin\frac{\theta_{p}-\theta_1}{2}
R_{\parallel}\left(\theta_{1}\right)X\left(k_{01}r_1, \theta_1-\theta_{p}\right), 
\end{eqnarray}
where 
\begin{eqnarray}\label{eq:20}
R_\parallel(\theta) = \frac{\frc{\dot{\varepsilon}_2}{\varepsilon_1}\cos\theta - \sqrt{n^2-\sin^2\theta}}
 {\frc{\dot{\varepsilon}_2}{\varepsilon_1}\cos\theta + \sqrt{n^2-\sin^2\theta}},\\
n = \frc{k_{02}}{k_{01}}=\sqrt{\frc{\dot{\varepsilon}_2\,\mu_2}{\varepsilon_1\,\mu_1}} 
\end{eqnarray}
are the Fresnel reflection factor and the index of refraction, respectively.

Note that the expression ($\ref{ter}$) corresponds to the space wave
\begin{equation}\label{eq:31}
E^{R}(r_1,\theta_1)=R_\parallel(\theta_1)\,E^{LOS}(r_1,\theta_1)
\end{equation}
 if one apply the large argument approximation for $X$ in (\ref{Lasimp}) provided
\begin{eqnarray} \label{Lcondit}
2k_{01}r_1\big|\sin\tfrac{(\theta_1-\theta_p)}{2}\big|^2> 1.
\end{eqnarray}

When the condition (\ref{Lcondit}) is ruled out, i.e.
\begin{equation}\label{con2}
p=k_{01}\rho\,\delta^2< 1,
\end{equation}
the solution (\ref{ter}) takes a novel form
 \begin{eqnarray}\label{eq:32}\nonumber
\vect{E}^{R}_{nsw}(\rho,x)=\vect{e}_{x}\delta \frac{pk_{01}^{3}}{2\varepsilon_{0}\varepsilon_{1}}\Big( e^{-\delta k_{01}\left(x+x_{0}+\delta\rho\right)}\\ \frac{e^{i\{k_{01}\left(\rho-\delta(x+x_0)\right)+\tfrac{\pi}2\}}}{2\sqrt{\pi k_{01}\rho}}+\tfrac{\delta}{\pi} e^{ik_{01}\rho}\Big)
\end{eqnarray}
near a planar interface, due to the asymptotics of the $X$ function in (\ref{asim}) for small argument approximation. 
 Here $p$ is the so-called Sommerfeld numerical distance \cite{Sommerfeld1926,Norton1937,Wait1998},  
\begin{eqnarray}\label{delta}
 \delta=\sqrt{\frc{\varepsilon_0\varepsilon_1\omega}{2\sigma}} 
\end{eqnarray} 
is the dimensionless parameter. The condition (\ref{con2}) can be obtained from (\ref{cond}), (\ref{th}) using simple substitutions:
 $$\kappa \to k_{01}\rho,\; \theta_1 \to \pi/2,\; |\alpha|=|\theta_1-\theta_p|\to \sqrt{2}~\delta.$$


Obviously, in case the numerical distance $p$ in (\ref{con2}) is greater than one, due to the increasing the distance $\rho$ (or the radiation frequency $ \omega$) and according to (\ref{Lcondit}), the kind of wave is converted to a space one.

 As we can see, a near-surface wave in (\ref{eq:32}) consists of a Zenneck surface wave and a uniform plane wave.
From the exponent of the abovementioned expression  (\ref{eq:32}), it follows that the phase velocity of the Zenneck wave is over the speed of light $c$
 \begin{equation} \label{vel}
v_f\simeq c\, \sqrt{1+\delta^2}
\end{equation}
 as well as a wavefront is slightly inclined forward. 

\subsection*{\textit{Lateral Waves}} 
 \label{later}                      

\begin{figure}[h]
\centering\includegraphics[width=3.in]{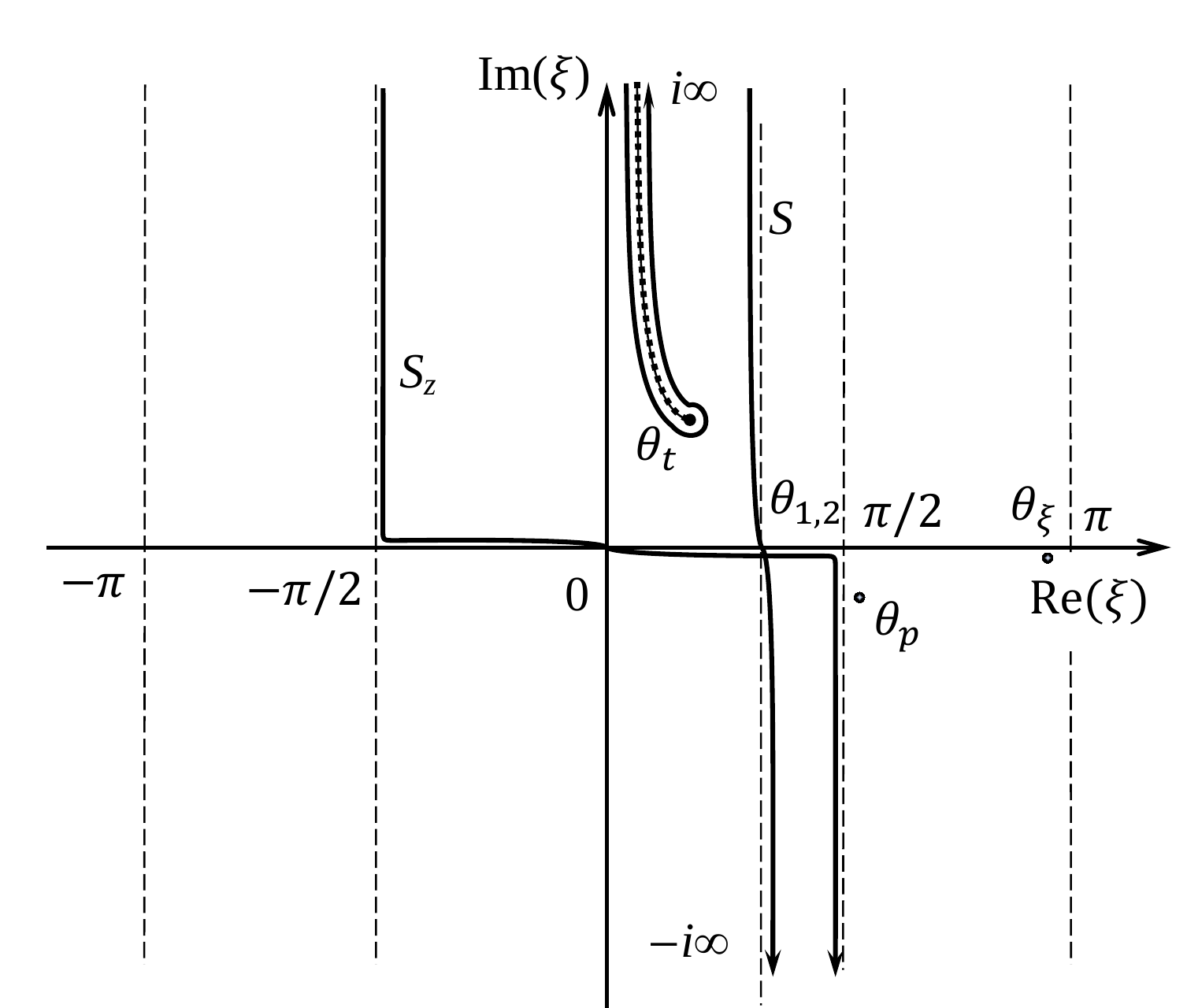}
\caption{The contours of integration.}\label{fig2}
\end{figure}

Let us introduce the following notations $k_\rho=k_{01}\sin\xi$,
$n=\sin\theta_t$  ($\theta_t=\arcsin n$) to calculating the integral (\ref{eq:ER}) along the banks of cut from the branch point $\theta_t$ (Fig.\ref{fig2}), which corresponds to lateral wave 
\begin{eqnarray}\label{eq:L1}\nonumber
\vect{E}^{R}_{Lat}(r_1,\theta_1)=-\frac{ipk_{01}^{3}}{8\pi\varepsilon_{0}\varepsilon_{1}}\sqrt{\frac{-2i}{\pi k_{01}\rho}}\int\limits_{\theta_t}^{i\infty}\vect{e}_{\theta}\left(\xi\right)
\sin^{\frac32}\xi\,\\
 e^{ik_{01}r_1\cos\left(\xi-\theta_1\right)}
\Big(\frac{\frac{\dot{\varepsilon}_2}{\varepsilon_1}\cos\xi-\sqrt{n^2-\sin^2\xi}}
{\frac{\dot{\varepsilon}_2}{\varepsilon_1}\cos\xi+\sqrt{n^2-\sin^2\xi}}-\nonumber\phantom{OO}\\
-\frac{\frac{\dot{\varepsilon}_2}{\varepsilon_1}\cos\xi+\sqrt{n^2-\sin^2\xi}}
{\frac{\dot{\varepsilon}_2}{\varepsilon_1}\cos\xi-\sqrt{n^2-\sin^2\xi}}\Big)d\xi=
\frac{ipk_{01}^{3}\dot{\varepsilon}_2}{2\pi\varepsilon_{0}\varepsilon_{1}^2}\sqrt{\frac{-2i}{\pi k_{01}\rho}}\nonumber\\
\int\limits_{\theta_t}^{i\infty}
\vect{e}_{\theta}(\xi)\,e^{ik_{01}r_1\cos
(\xi-\theta_1)}
\frac{\sin^{\frac32}\xi\,\cos\xi\,\sqrt{n^2-\sin^2\xi}}
{(\dot{\varepsilon}_2/\varepsilon_1)^2\cos^2\xi-n^2+\sin^2\xi}d\xi,
\end{eqnarray}
where $\theta_t$ is the angle of total reflection, $$r_1=\sqrt{\rho^2+(x+x_0)^2},\; \tan\theta_1=\rho/(x+x_0).$$

In order to calculate the integral, it would be useful to take advantage of the integral representation of the parabolic cylinder (Weber) function \cite{abramowitz+stegun} as the etalon integral
\begin{equation}\label{eq:L2}
U(1,z)=\frac{e^{-z^2/4}}{\Gamma(\frac32)}\int\limits_0^\infty
e^{-zs-s^2/2}\sqrt{s}\,ds,
\end{equation}
where $\Gamma$ is the gamma function.

Introducing a new variable of integration $s=\alpha(\xi-\theta_t)$ with the following notations
\begin{equation} \label{eq:L3}
 \alpha^2=ik_{01}\,r_1\cos\theta^*,\quad \theta^*=\theta_t-\theta_1,
\end{equation}
  and taking into account the series expansion in the neighbourhood of the point $s=0$ of the functions:
\begin{eqnarray*}
\sqrt{n^2-\sin^2(s/\alpha+\theta_t)}= \sqrt{-\frac2{\alpha}\cos{\theta_t}\sin \theta_t}~\sqrt{s}+\dots,\\
\cos(s/\alpha+\theta^*)= \cos\theta^*-\frac{\sin\theta^*}{\alpha}\,s- \frac{\cos\theta^*}{2\alpha^2}\,s^2+\dots,
\end{eqnarray*}
it is easy to find the main contribution to the integral (\ref{eq:L1}) using SPM with help of the Weber function $U$ in (\ref{eq:L2})
\begin{eqnarray}\label{eq:L4}
\vect{E}^{R}_{Lat}(r_1,\theta_1)= -\vect{e}_{\theta}(\theta_t)
\frac{k_{01}\,p}{2\pi\varepsilon_0\dot{\varepsilon}_2}\frac{\sin^2\theta_t}{\sqrt{\rho\cos\theta_t}}\left(\frac{i k_{01}}{r_1\cos\theta^*}\right)^{\frac34}\nonumber\\
 e^{ik_{01}r_1\cos\theta^*\left(4+\tan^2\theta^*\right)/4}\, 
U(1,\alpha \tan\theta^*).
\end{eqnarray}

\begin{figure}[h]
\centering\includegraphics[width=3.in]{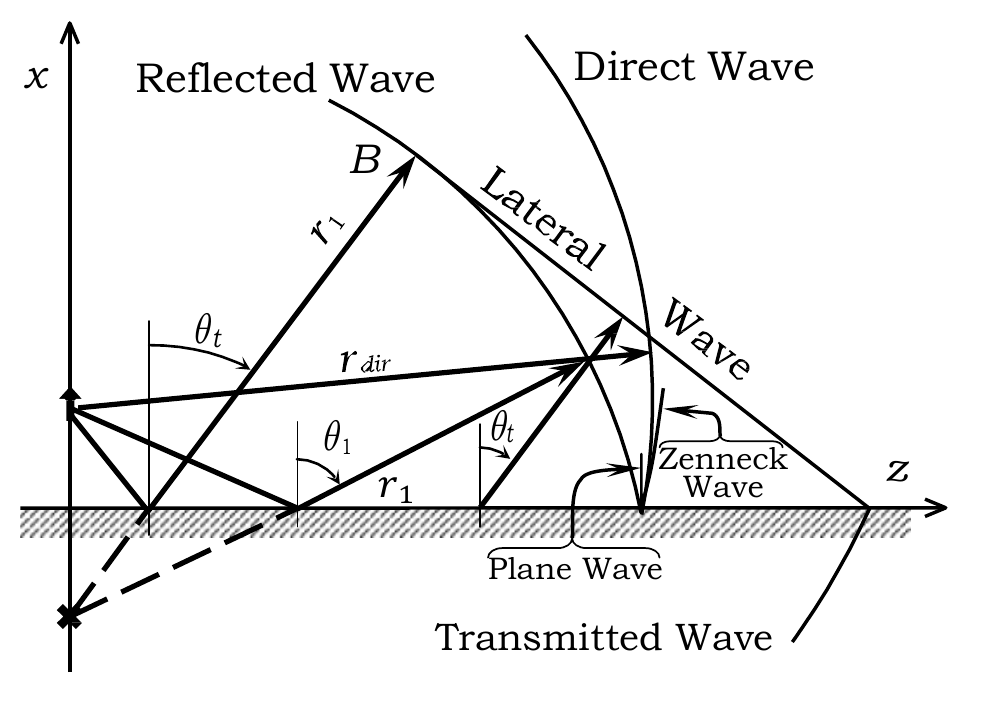}
\caption{Location of wave fronts.}\label{fig3}
\end{figure}

Taking into consideration  the asymptotic formula of the Weber function
$$
U(1,z)\sim z^{-\frac32}e^{-\frac14 z^2} \quad \text{at}\quad |z|>1,
$$
we have the asymptotic behaviour of (\ref{eq:L4}) (at observation point  $A$ in Fig.\ref{fig3})
\begin{eqnarray}\label{eq:L5}\nonumber
\vect{E}^{R}_{Lat}(r_1,\theta_1)&\simeq& \vect{e}_{\theta}(\theta_t)
\frac{k_{01}p}{2\pi\varepsilon_0\varepsilon_1}\frac{\mu_2}{\mu_1}\\
&&\frac{e^{ik_{01}
(n\rho+(x+x_0)\sqrt{1-n^2})}}{(n^2-1)(\tan \theta_t\cot\theta_1-1)^{\frac32}\,\rho^2}
\end{eqnarray}
under given condition

$k_{01}r_1|\sin\theta^*|^2> |\cos\theta^*|$\; \text{or} \; $k_{01}r_1|\theta_1-\theta_t|^2>1$.

Thus, we have the same expression for the lateral waves obtained by L.M. Brekhovskikh \cite{Brekh} in the case of (\ref{eq:L5}).
As can be seen from the formula (\ref{eq:L5}), a wave front is linear in the plane ($\rho$, $x$)
(Fig.\ref{fig3}, $B$ represents an observation point). Therefore, the lateral wave is conical.

It is to be noted that the above-mentioned  outcome (\ref{eq:L5}) is unusable if the angle of incidence $\theta_1$ is close to the angle 
of total reflection $\theta_t$.
Thus, using the asymptotic form of the Weber function at $z<1$ 
$$
U(1,z)\sim \frac{\sqrt{\pi}}{2^{3/4}\Gamma(\frac54)}(1-z),
$$
 we obtain the asymptotic behaviour of the lateral wave near the critical angle for total-internal reflection  $\theta_1 \to \theta_t$ ( $\dot{\varepsilon}_2=\varepsilon_2$)
\begin{eqnarray}\label{eq:L6}\nonumber
\vect{E}^{R}_{Lat}(r_1,\theta_{t})\simeq -i\vect{e}_{\theta}(\theta_t)
\frac{k_{01}^2p}{\varepsilon_0\varepsilon_1\Gamma(\frac14)}\phantom{OOOOOOOooooo}\\
\frac{\mu_2}{\mu_1}\left(\frac{2}{k_{01}r_1(1-n^2)}\right)^\frac14
\frac{e^{i(k_{01}r_1-\pi/8)}}{r_1\sqrt{\pi n}}.
\end{eqnarray}

In case of  $\mu_2=\mu_1 = 1$, $n^2=\varepsilon_2/\varepsilon_1<1$ we have
\begin{equation}\label{eq:L7}
\vect{E}^{R}_{Lat}(r_1,\theta_t)\simeq -\vect{e}_{\theta}(\theta_t)
 \frac{0.328}{\varepsilon_0\varepsilon_1}
\frac{p\,k_{01}^2}{(k_{01}\,r_1)^\frac14}
\frac{e^{i(k_{01}\,r_1+\pi/8)}}{r_1 n \sqrt{\pi}}
\end{equation}
or
\begin{equation}\label{eq:L8}
\vect{E}^{R}_{Lat}(r_1,\theta_t)= \vect{e}_{\theta}(\theta_t)\frac{2.325\, e^{i \pi/8}}{n^2(k_{01}\,r_1)^{\frac14}}\,E^{LOS}(r_1,\theta_t). 
\end{equation}

It is obvious that the intensity of the lateral wave at the observation point $B$ (see Fig.\ref{fig3}) is greater than at the observation point $A$, as can be seen from  (\ref{eq:L5}) and (\ref{eq:L6}) or (\ref{eq:L8}).

\subsection*{\textit{Transmitted waves}}

To calculate the integrals in (\ref{eq:11}, \ref{eq:12}) by  SPM, we need to find the saddle point of the exponential function
\begin{equation}\label{p0}
 e^{i\varPhi}=e^{i\left(\sqrt{k_{01}^2-\rho^2} x_0-
 \sqrt{k_{02}^2-\rho^2} x+\rho k_\rho\right)}
\end{equation}
on the complex plane of the new integration variable $\xi$ using the transformation, which we known as
\begin{equation*}\label{p1}
k_\rho=k_{02}\sin\xi.
\end{equation*}

\begin{figure}[h]
\centering\includegraphics[width=3.2 in]{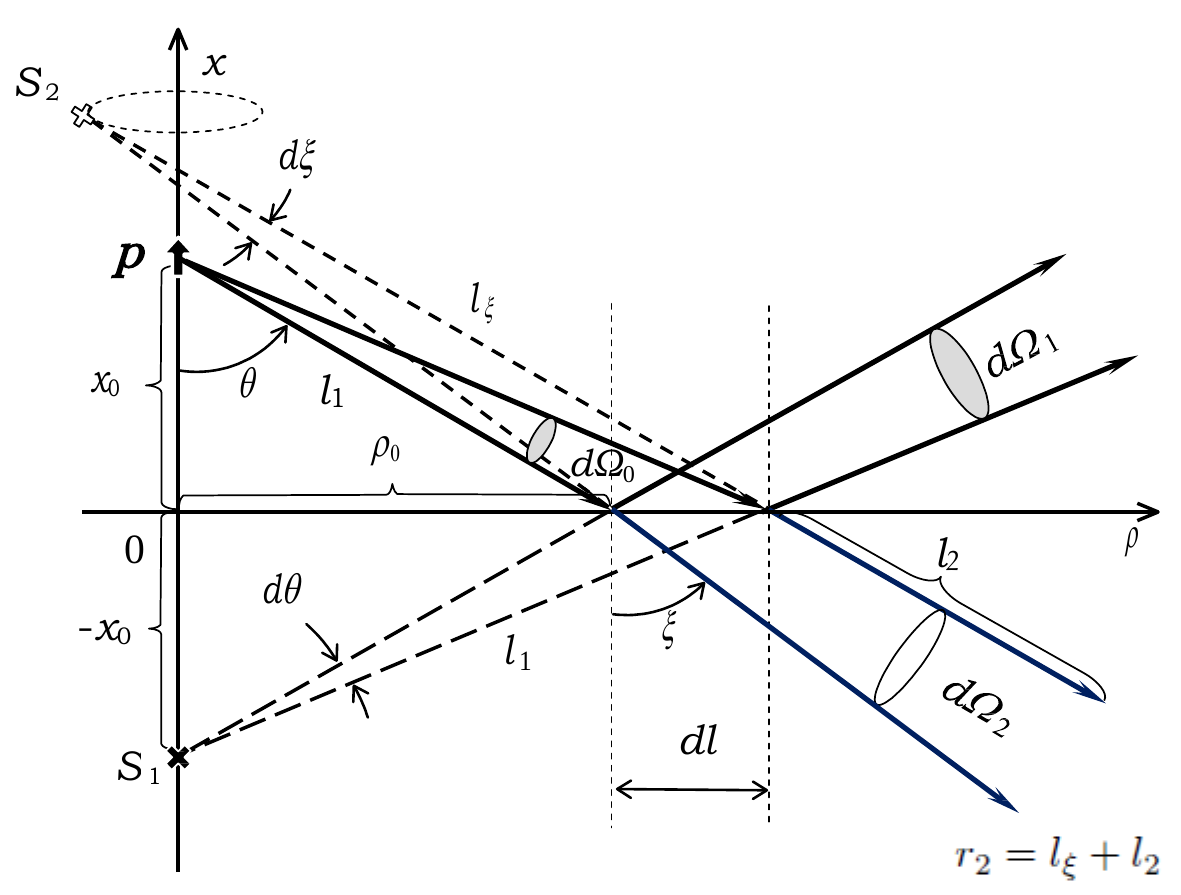}
\caption{Ray reflection and refraction inside a thin solid angle $d\Omega_1$.}\label{fig4}
\end{figure}

The phase function in $\xi$ can be represented as
\begin{eqnarray}\label{Phaz}
\varPhi(\xi)=k_{02}\rho\sin\xi-k_{02}x\cos\xi+k_{01}x_0\cos\theta(\xi).
\end{eqnarray}
It bears mentioning that the notation entered herein (\ref{Phaz})
\begin{eqnarray}\label{phazz}
\cos\theta(\xi)=\sqrt{1-(n\sin\xi)^2},  
\end{eqnarray}
as is easily guess, corresponds to the Snell's law of geometric optics (see Appendix \ref{Phi}, (\ref{spxi}))
\begin{eqnarray}\label{Snell}
\frac{\sin\theta}{\sin\xi}=\frac{k_{02}}{k_{01}},
\end{eqnarray}
 where $\xi$ corresponds to the ray refraction angle in the second medium (Fig.\ref{fig4}), and $\theta$ is the angle of incidence.

From Snell's law follows a useful relation hereinafter
\begin{eqnarray}\label{pk}
K(\xi)\equiv\frac{d\theta}{d\xi}=\frac{\tan\theta}{\tan\xi}=\frac1n\frac{d\Omega_1}{d\Omega_2},
\end{eqnarray}
the meaning of which is the relative change of the solid angle $d\Omega_1$ of a thin ray tube when transmitted into a second medium with a solid angle $d\Omega_2$.

We will assume the permittivity of the second medium to be real ($\dot{\varepsilon_2}=\varepsilon_2$) since it is not an easy issue to determine the saddle point.

Owing the approximating phase function in (\ref{asPhi}), the integral in (\ref{eq:11}) is represented as
\begin{eqnarray}\label{ETi}\nonumber
\vect{E}^T(r_2,\theta_2)=\frac{pk_{02}^2k_{01}}{8\pi i\varepsilon_0\varepsilon_1}
\sqrt{\frac{-2i}{\pi k_{02}\rho}}
e^{i\left(\varPhi(\theta_2)-k_{02}\,r_2\right)}\\
\int_{S_z}\vect{e}_{\theta_2}(\xi)
e^{ik_{02}r_2\cos(\xi-\theta_2)}
K(\xi)T_{\parallel}(\xi) 
\sin^{\frac32}\xi\, d\xi,
\end{eqnarray}
where\begin{eqnarray} \label{polt}
T_\parallel(\xi)=\frac{2\varepsilon_1n\cos \theta(\xi)}{\varepsilon_1n\cos\xi+\varepsilon_2\cos \theta(\xi)},
\end{eqnarray}
the saddle point $\theta_2$ (\ref{spxi}) is defined from the vanishing of the first derivative of the phase function
$\partial \varPhi/\partial\xi=0$ (Appendix \ref{Phi}, (\ref{pp4})), the expression
\begin{eqnarray} \label{eth2} 
\varPhi(\theta_2)-k_{02}\,r_2=
k_{01}\frac{x_0}{\cos\theta_1}(1-K^2) 
\end{eqnarray}
is followed from (\ref{Phi2}), (\ref{llk}) with $K=\frc{-x \rho_0}{x_0(\rho-\rho_0)}$.

By deforming the integration contour $S_z$ to $S$ (Fig. \ref{fig2}) and using the etalon Integral (\ref{xi}), we obtain an asymptotic formula from (\ref{ETi}) in the short-wave approximation   
\begin{eqnarray} \label{EtX}
\vect{E}^T(r_2,\theta_2)=\vect e_{\theta_2}\frac{pk_{02}^2k_{01}}{2\varepsilon_0\varepsilon_1}
e^{i\big(\varPhi(\theta_2)-k_{02}\,r_2[1-\cos(\theta_2-\xi_p)]\big)}\nonumber\\
K\,\sqrt{\frac{-2i}{\pi k_{02}\rho}}T_{\parallel}(\theta_2) 
\sin^{\frac32}\theta_2\,\sin\tfrac{\theta_2-\xi_p}{2}\,
X(r_2 k_{02},\theta_2-\xi_p),
\end{eqnarray}
 similar to $\vect E^R$.
 
The space wave field in the far zone
\begin{equation}\label{ET}
\vect{E}^T(r_2,\theta_2)=-\vect e_{\theta_2}
\frac{pk_{01}^2}{4\pi \varepsilon_0\varepsilon_1}
\frac{e^{i\varPhi(\theta_2)}}{r_2}\sin\theta_2\,T_\parallel(\theta_2)\,\frac{d\Omega_1}{d\Omega_2}
\end{equation}
follows from (\ref{EtX}), bearing in mind that $\rho \sim r_2\sin\theta_2$ as well as the asymptotics of the special function $X$ in (\ref{Lasimp}) with a condition
\begin{equation} \label{space} 
 2k_{02}r_2\left|\sin\tfrac{\theta_2-\xi_p}2\right|^2>1, 
\end{equation} 
where $d\Omega_1/d\Omega_2=n^2\cos\theta_2/\cos\theta_1$ is the relative change of the solid angle of a conical ray tube (\ref{pk}).

\section{Discussion}

It bears mentioning that the obtained asymptotic solutions for reflected (\ref{eq:31}) and transmitted (\ref{ET}) space waves obey the local law of conservation of energy flux provided $k_{01}$ and $k_{02}$ are real, as well as $\theta_1<\theta_t$.

Let us choose a thin conical tube, the lateral surface of which is formed by rays outgoing from the point dipole (Fig. \ref{fig4}).  The time-averaged value of the energy flux $dW$ inside the solid angle $d\Omega$ of the ray tube defined as the product of the cross-section of a cone and an average absolute value of the  Poynting vector
\begin{equation}\label{w}
dW=\frac12 E_\theta \bar{H}_\alpha r^2d\Omega, 
\end{equation}
where a bar over $H$ denotes complex conjugation.

Taking into account the fields (\ref{Alos}) and (\ref{eq:18}) in (\ref{w}), we find the energy flux of the source
\begin{equation}\label{W0}
dW_0=\frac12 A^2\sin^2\theta_1Z^{-1}_1d\Omega_0, \quad A=\frac{pk_{01}^2}{4\pi\varepsilon_0\varepsilon_1}
\end{equation}
inside the solid angle $d\Omega_0$.
Similarly, we define the energy fluxes for reflected $dW_1$ and transmitted waves $dW_2$ inside the solid angles
 $d\Omega_1$ and $d\Omega_2$, according to (\ref{eq:31}) and (\ref{ET}):
\begin{eqnarray}
&{}dW_1=\frac12A^2\sin^2\theta_1R_\parallel^2
Z^{-1}_1d\Omega_1, \label{W1}\\ &{}dW_2=\frac12A^2\sin^2\theta_2T_\parallel^2
Z^{-1}_2\Big(\frac{d\Omega_1}{d\Omega_2}\Big)^2d\Omega_2,\label{W2}
\end{eqnarray}
where 
\begin{equation*}
R_\parallel= \frac{Z_1\cos\theta_1-Z_2\cos\theta_2}{Z_1\cos\theta_1+Z_2\cos\theta_2}, \quad T_\parallel=\frac{2Z_2\cos\theta_1}{Z_1\cos\theta_1+Z_2\cos\theta_2}
\end{equation*}
are Fresnel's coefficients of reflection  (\ref{eq:20}) and transmission (\ref{polt}).
We take into account that the solid angles of ray tubes in the first medium are the same $(d\Omega_0=d\Omega_1)$, here $\theta_1$ and $\theta_2$ are the angles of flux reflection and transmission.

Now we can make ascertain that the fields of reflected and transmitted waves inside the corresponding ray tubes satisfy the law of conservation of energy flux.
\begin{equation}\label{w123}
dW_0=dW_1+dW_2.
\end{equation}
Indeed, we obtain the identity from (\ref{w123})
\begin{equation} \label{RT2}
1-R^2_\parallel\equiv T_\parallel^2\,\frac{Z_1\,\cos\theta_2}{Z_2\,\cos\theta_1}
\end{equation}
counting the above-mentioned expressions in (\ref{W0},  \ref{W1}, \ref{W2}) 

As can be seen from (\ref{ET}) and (\ref{eq:31}), expressions for the reflected and transmitted space waves wear defined as the product of the source field (\ref{Alos}) and the Fresnel reflection $R_\parallel$ or transmission $T_\parallel$ coefficient, respectively.

 This form of solution can be easily explained from the standpoint of the geometric optics since a spherical wavefront inside a thin ray tube is considered as a plane front. 
Thus, a part of the problem of finding the space wave fields reduced to solving the Fresnel problem of the transmission and reflection of plane waves at a planar interface of two media. 
 
Indeed, we find the solution by multiplying the Fresnel reflection coefficient on the expression for the field of a virtual source $S_1$, which is a mirror image of a dipole in the second medium.
 The transmitted wavefield (\ref{ET}) from the virtual image $S_2$ located in the first medium is defined similarly, taking account of the Fresnel transmission coefficient  $T_\parallel$ as well as the relative change of the solid angle $d\Omega_1/d\Omega_2$. 
 
It is interesting to note that the virtual image $S_2$ of a transmitted wave has a ring shape (see Fig. \ref{fig4}), as opposed to the point image $S_1$.

Now let's discuss the controversial issue of the existence of surface waves propagating along the planar interface between two media.
 
From one side, we have the asymptotic solution in the form of a fast Zenneck surface wave in (\ref{eq:32}) only in the neighbourhood of the unit radius of the so-called Sommerfeld numerical distance $p$, which follows from the more general solution (\ref{ter}) covering all near-surface waves.
 On the other side, the residue term is not the contribution from the integral (\ref{eq:ER}) at a pole of the function $R_\parallel$ in (\ref{th}) corresponding to the true (independent) surface wave, since the Sommerfeld integration contour fails to capture the pole $\theta_p$ when it deforms to the path of steepest descent (see Fig.2). Therefore, it seems that such a 'ghost' wave is a surface wave in form but not in content, since it ceases to exist outside the limit of a unit numerical distance.  This dualism in solving the Sommerfeld problem seems to be the main reason for the long-running dispute about the reality of the surface wave in the scientific world.

Thus, we can conclude that the discovered wave in (\ref{eq:32}) is not a true surface wave but nevertheless is generally referred to as the surface wave. Rather it is a pseudo-surface wave since the surface wave (\ref{eq:31}) transits into the space wave at some distance due to the property of the function $X$ when the pole $\theta_p$ and the saddle point $\theta_1$ are very close. Recall that the fast Zenneck wave is accompanied by a uniform plane wave, although the first one predominates, as seen from (\ref{eq:32}).      

 It is important to note that the exact solution (\ref{eq:ER}) of the boundary value problem contains a solution (\ref{eq:L4}) as the lateral wave in addition to the above space and surface waves.  
 
The lateral wave solution derives from (\ref{eq:ER}) by deforming the Sommerfeld integration contour $S_z$ to $S$, passed along the banks of the cut in (\ref{eq:L1}), which begins from the branch point $\theta_t$ of the function $R_\parallel$ upward in the direction of the imaginary axis $i\infty$ (see Fig.\ref{fig2}) of the complex plane $\xi$. 
  
It is evident that the angle of total internal reflection ($\theta_t$) is complex in a less optically dense medium  $|\sin\theta_t|=|n|>1$, therefore the lateral wave field is exponentially decayed in the direction of the $x$-axis, as the following exponential factor evaluation occurs in
 (\ref{eq:L5}) \cite{Leontovic1946}
\begin{equation} \label{exp}
 e^{i k_{01}(n\rho+(x_0+x)\sqrt{1-n^2})}\sim 
 \begin{cases}
e^{-k_{01}\sqrt{\varepsilon_2/\varepsilon_1}(x_0+x)} \quad  (\sigma=0),\\
e^{-\frac{k_{01}}{2\delta}(x_0+x+\rho)}  \quad (\delta < 1).
 \end{cases}
 \end{equation}

In the case of the source is in a more optically dense medium ($|\dot{\varepsilon}_2| < \varepsilon_1$ or $n=\sin\theta_t <1$), the lateral wave amplitude decreases with the distance as $1 / \rho^2$ (\ref{eq:L5}) and increases with the distance from the interface by increasing the factor $(\tan\theta_t\cot\theta_1-1)^{-3/2}$,  provided $\sqrt{k_{01}r_1}\,|\theta_1-\theta_t|>1$.
 In the vicinity of the observation angle $\theta_1=\theta_t$ the formula is unjust, it should be substituted by the formula (\ref{eq:L7}), which follows from (\ref{eq:L4}) and is expressed by $E^{LOS}$ with a constant unit vector $\vect{e}_{\theta}(\theta_t)$ in (\ref{eq:L8}).
 
The appearance of the lateral wave can be interpreted as a reradiation of the non-uniform waves, traveling along the plane interface into the outdoor space at an angle of total internal reflection $\theta_t$ according to Snell's law.

 Hence it follows that a lateral wave can exist provided $\rho > x_0 \sqrt{\varepsilon_2/\varepsilon_1}$, which is equivalent to the condition $\theta_1 > \theta_t$, which follows from the Weber function domain.  

Thus using the etalon integral (\ref{eq:L2}) expressed by the Weber function and taking into account the mutual influence of the pole and the branch point in the integrand, it is possible to obtain a solution in the form of a lateral wave  regular for any values of observation angles.

\section{Conclusions}

In contrast to the traditional method for solving the Sommerfeld problem, the solution deduced concerning the densities of the currents in the Fourier transform domain induced at the planar interface of two media, without using a vector potential.  Therefore the method can be extended for anisotropic media \cite{sautbek10}.
  
In this paper, the regular for any observation angles asymptotic expressions for the fields of the reflected, transmitted, and lateral waves wear obtained due to the modified SPM with an etalon integral for the integration of the Sommerfeld contour integrals.  

The solution of the Sommerfeld problem on the transmission of a wave into a second medium, which is more intricate, was reduced to the considered problem of a reflected wave due to the proposed simple method of finding the saddle point $\theta_2$ and the approximating phase function $\varPhi(\xi)$. 

Although the asymptotic solutions for the reflected and transmitted waves should coincide in form, the latter differs only in the factor $d\Omega_1/d\Omega_2$ in (\ref{pk}), which corresponds to the ratio of the solid angles  of the conical ray tubes at the interface of the media.

Based on the interpretation of the physical meaning of the obtained solutions for reflected and transmitted space waves, a simple constructing technique for the corresponding solutions is proposed with the help of the Fresnel formulae for a plane wave.
 
It bears mentioning that the obtained asymptotic solutions for reflected (\ref{eq:31}) and transmitted (\ref{ET}) space waves obey the law of conservation of energy flux of the incident wave (\ref{w123}) and reduce to the identity (\ref{RT2}) for the Fresnel coefficients $R_\parallel$ and $T_\parallel$ provided $k_{01}$ and $k_{02}$ are real. 

 The outcomes could be generalized for smooth curved media interfaces since the physical meaning of the obtained asymptotic expressions for reflected and transmitted waves is easily explained using the ray tubes technique, inside the wavefront is considered locally plane.

It has been shown the near-surface waves decompose into a uniform plane wave as well as Zenneck wave, which phase velocity over the light speed and its wavefront slightly tilted forward.

In the second medium, there are no surface waves because the Sommerfeld integration path $S_z$ does not capture the pole $\theta_p$ when it deforms to the steepest descent path $S$, as well as, it is quite far from the pole $\theta_p$.

The solution in the lateral waveform, regular for any observing angle, corresponds to the integration contour along the banks of the cut outgoing from the branch point upward along the imaginary axis of the complex plane. Here Weber's function is as an etalon integral \cite{abramowitz+stegun}. One of the asymptotics of the solution matches the result of the work of L.M. Brekhovskikh \cite{Brekh}.
 
In the case of  $n>1$, the wavefront propagates normal to the plane interface, and a field amplitude exponentially decays from the interface plane.
 In the case of the source is in a more optically dense medium $n<1$,  the lateral wave amplitude decreases inversely as the distance squared to the source, and the wavefront is conical. 
 
\appendix

\numberwithin{equation}{section}

\refstepcounter{equation}
\section{Special Function $X(\kappa,\alpha)$}

To evaluate the short-wave asymptotic integrals by the saddle point method, if a pole close to the saddle point, it is advantageous to use the etalon integral $S$  \cite{sautbek17,sautbek18}, 
which reduces to Fresnel integrals or the error function integral $\textrm{erf}(z)$  \cite{abramowitz+stegun}
 \begin{eqnarray}\nonumber
X(\kappa,\alpha)=\frac{1}{4\pi i}\int_Se^{i\kappa(\cos\xi-\cos\alpha)}
\frac{d\xi}{\sin\frac{\xi+\alpha}{2}}=\\
=\frac{e^{-i\frac\pi 4}}{\sqrt{2\pi}}  
\int^{2\sqrt{\kappa}~\sin\frac\alpha 2}_{\infty\sin\frac\alpha2}e^{i\frac{t^2}{2}}~dt= \nonumber   \\
-\frac12 ~{\textrm{sgn}\big(\textrm{Re}(\alpha)\big)} +
\frac12~{\textrm{erf}\left(\sqrt{-i2k}~\sin \frac\alpha2\right)}. \quad \label{xi} 
\end{eqnarray}

The function $X$ has the following asymptotics:
\begin{eqnarray}\label{Lasimp}
X\left(\kappa,\alpha\right)\simeq -\sqrt{\frac{i}{2\pi \kappa}}\frac{e^{i\kappa(1-\cos\alpha)}}{2\sin\frac{\alpha}{2}},
\end{eqnarray}
at large values of the argument
\begin{eqnarray}\label{Con}
 2\kappa|\sin\tfrac\alpha 2|^2> 1,
 \end{eqnarray}  
as well as
  \begin{eqnarray}\label{asim}	\nonumber
X(\kappa,\alpha)+\frac12\textrm{sgn}
\big(\textrm{Re}(\alpha)\big)\simeq \phantom{FFFFFFFFFF}\\ 
\simeq\sqrt{\frac{2\kappa}{i\pi}}\sin\tfrac\alpha2~e^{i\kappa(1-\cos\alpha)}
\simeq \sqrt{\frac{\kappa}{2\pi i}}~\alpha 
\end{eqnarray}
at small values of the argument
\begin{eqnarray}\label{cond}
 \kappa|\alpha|^2/2<1.
 \end{eqnarray}

 \section{Evaluation of the poles of the reflection $R_\parallel$ and transmission $T_\parallel$ coefficients} 

By transforming the integration variable for the first medium
\begin{eqnarray}\nonumber
k_\rho=k_{01}\sin\theta, 
\end{eqnarray}
we find the root $\theta_p$ of the equation
\begin{eqnarray} \label{eqR}
\varepsilon_1\sqrt{k^2_{02}-k^2_{01}\sin^2\theta}+\varepsilon_2k_{01}\cos\theta=0,
\end{eqnarray}
which is the pole of a reflection coefficient $R_\parallel$ (\ref{eq:20}) in the complex plane $\theta$ 
\begin{eqnarray}\label{sil}
\cos\theta_p=-\sqrt{\left(\frc{k^2_{02}}{k^2_{01}}-1\right)\Big{/}\left(\frc{\dot{\varepsilon}_2^2}{\varepsilon_1^2}-1\right)}.
\end{eqnarray}
It should be noted that for the correct choice of the square root sign (\ref{sil}), there is a good reason to solve the equation (\ref{eqR}) concerning $\sin^2\theta$, then obtain (\ref{sil}).

For nonmagnetic media ($\mu_1=\mu_2=1$), the following estimation is valid for the pole of the reflection coefficient $R_\parallel$
\begin{eqnarray}\label{aspole}
\theta_p\simeq\frac{\pi}{2}+\sqrt{\frac{\varepsilon_1}{(\varepsilon_1+\dot{\varepsilon}_2}}\simeq \frac{\pi}{2}+\sqrt{\frac{\varepsilon_1}{\dot{\varepsilon}_2}}\,\Big(1-\frac{\varepsilon_1}{\dot{\varepsilon}_2}\Big),
\end{eqnarray}
which follows from (\ref{sil}), i.e.
\begin{eqnarray}\label{aspol}
\cos\theta_p=-\sqrt{\frac{\varepsilon_1}{(\varepsilon_1+
\dot{\varepsilon}_2)}}.
\end{eqnarray}

In the case accounting of conductivity of the second medium and also provided that $\delta <1$, we can obtain the approximate value of the pole for the reflected wave
\begin{eqnarray}\label{th}
\theta_p=\pi/2+(1-i)\delta+(1+i)(1+\varepsilon_2/\varepsilon_1)\delta^3.
\end{eqnarray}

Now, similarly to (\ref{sil}), we can find the pole of the transmission coefficient $T_\parallel$ in the complex plane
 $\xi$ 
\begin{eqnarray}\label{xipol}
\cos\xi_p=-\sqrt{\left(\frc{k^2_{01}}{k^2_{02}}-1\right)\Big{/}\left(\frc{\varepsilon_1^2}{\dot{\varepsilon}_2^2}-1\right)},
\end{eqnarray}
which corresponds to the transformation of the integration variable $k_\rho=k_{02}\sin\xi$ for the second medium issue and is the root of the equation
\begin{eqnarray} \label{eqT}
\varepsilon_2\sqrt{k^2_{01}-k^2_{02}\sin^2\xi}+\varepsilon_1k_{02}\cos\xi=0.
\end{eqnarray}
In the case of non-magnetic media, the expression for the pole $\xi_p$ in (\ref{xipol}) is written in the form 
\begin{eqnarray}\label{xipmu}
\cos\xi_p=-\sqrt{\frc{\dot{\varepsilon}_2}{(\varepsilon_1+\varepsilon_2)}}.
\end{eqnarray}

It is seen the following estimation of the pole occurs
\begin{eqnarray}\label{isol}
\xi_p\sim \pi-\sqrt{\varepsilon_1/\varepsilon_2} \quad (\varepsilon_2>\varepsilon_1, \;  \dot{\varepsilon}_2=\varepsilon_2 ).
\end{eqnarray}

 \section{Determination of the branch point. The phase function of the transmitted wave} \label{Phi}

Taking the first derivative of the phase function (\ref{Phaz}), we obtain the equation for determining the branch point
\begin{eqnarray}\label{pp4}
\varPhi'_\xi(\xi)=k_{02}(\rho \cos\xi+x\sin\xi-x_0\cos\xi\tan\theta)=0.
\end{eqnarray}

By inserting the parameter $\rho_0$ in (\ref{pp4}), which defines the reflection point coordinate of the ray from the interface ($\rho=\rho_0$ at $x=0$) in Fig.\ref{fig4}, as well as writing the equation in the form
\begin{eqnarray}\label{p3}
\rho_0 =x_0\tan\theta=\rho+x\tan\xi,
\end{eqnarray}
 we can easily find the ray incidence angle in the first medium
\begin{eqnarray}\label{p5}
\theta_1=\arctan (\rho_0/x_0),
\end{eqnarray} 
 which is the branch point in the expressions (\ref{ter}) and (\ref{eq:31}) for the reflected wave that important to note.

Similarly,  taking into account the expression (\ref{phazz}), through the parameter $\rho_0$, we find the root of the equation (\ref{pp4}) or (\ref{p3}) 
\begin{eqnarray}\label{spxi}
\theta_2=\arcsin\big(\tfrac1n\sin\theta_1 \big)=\arcsin\Bigg(\frac{k_{01}\rho_0}{k_{02}\sqrt{\rho^2_0 +x^2_0}}\Bigg),
\end{eqnarray} 
which is the branch point for the second medium and corresponds to the angle of ray refraction(\ref{Snell}) transmitting into the second medium.
As we can see, the value of the branch point can be obtained directly from Snell's law (\ref{Snell}).

The coefficient $K$ in (\ref{pk}) derives by using the formulae (\ref{Snell}) and $d\Omega_1=\sin\theta\, d\theta\, d\alpha$ for an elementary solid angle both in the first medium and in the second $d\Omega_2=\sin\xi\, d\xi\, d\alpha$.

Since the ray trajectory determines according to Snell's law, the coordinates of the observation point $(\rho, x)$ have to select via the parameter $\rho_0$ according to the expression 
\begin{eqnarray}\label{p6}
\rho=\rho_0-x \tan\theta_2,
\end{eqnarray}
which follows from (\ref{p3}).

Now from (\ref{pp4}), using (\ref{Snell}) and making geometric calculations (Fig. \ref{fig4}), we calculate the second derivative at the branch point
\begin{eqnarray} \label{sht} \nonumber
-\frac{\varPhi^{''}_\xi (\theta_2)}{k_{02}}=\rho\sin\theta_2-x\cos\theta_2-x_0 \phantom{FFFFFFFF}\\
\Big(\sin\theta_2\tan\theta_1-\frac{\cos^2\theta_2}{\cos^2\theta_1}\frac{\tan\theta_1}{\sin\theta_2}\Big).
\end{eqnarray}

By introducing from geometric considerations the following notations in (\ref{sht})
\begin{eqnarray}\label{p9}
l_2\equiv \frac{-x}{\cos\theta_2}, \;  
l_{\xi}\equiv \frac{k_{01}}{k_{02}}K^2l_1, \; l_1 \equiv  \frac{x_0}{\cos\theta_1}, 
\end{eqnarray}
where $l_1$ and $l_2$ are the path of the ray in the first and second media and equal to, respectively
\begin{eqnarray}\label{pl0}
l_2=(\rho-\rho_0)\sin\theta_2-x\cos\theta_2,\quad 
l_1=\sqrt{x_0^2+\rho_0^2},
\end{eqnarray}
the abovementioned expression can be written
 \begin{eqnarray}\label{llk}
-\varPhi^{''}_\xi (\theta_2) = k_{02}(l_\xi +l_2)\equiv k_{02} r_2.
\end{eqnarray}
The geometric meaning of the abovementioned values is that $r_2$ is the distance from the observation point to the imaginary source,
$l_2$ is a part of the line segment $r_2$ in the second medium, $l_\xi$ is a part in the first medium (see Fig. \ref{fig4}).

Based on geometric reasons, it is interesting to note that the expression for $l_\xi$ in (\ref{p9}) can be obtained from the equality of the general hypotenuse of triangles $dl$, lying on a planar interface the legs of which are formed by the rays of the ray tube at reflection and transmission (Fig.\ref{fig4})
\begin{eqnarray}
dl=\frc{l_\xi\, d\xi}{\cos\xi}=\frc{l_1\, d\theta}{\cos\theta}.
\end{eqnarray}

To preserve the unity of the integration technique, it is good practice to decompose the phase function by cosine in the vicinity of the branch point, taking into account its second derivative
\begin{eqnarray}\label{}
\varPhi(\xi)=\varPhi(\theta_2)+\varPhi^{''}_\xi(\theta_2)\big(1-\cos(\xi-\theta_2)\big)+ \cdots
\end{eqnarray} 
the phase function approximately can be represented
\begin{eqnarray}\label{asPhi}
\varPhi(\xi)\cong\varPhi(\theta_2)-k_{02}r_2+k_{02}r_2\cos(\xi-\theta_2),
\end{eqnarray}
where
\begin{eqnarray}\label{Phi2}
\varPhi(\theta_2)=k_{01}l_1+k_{02}l_2=k_{01}\frac{x_0}{\cos\theta_1}+k_{02}\frac{-x}{\cos\theta_2}.
\end{eqnarray}

\section*{References} 

 \bibliography{TAP20} 
\end{document}